\documentclass[prl,twocolumn, amsmath,amssymb,superscriptaddress]{revtex4-2}
\usepackage{graphicx,subfigure,amsmath,bm,colordvi,times}
\usepackage{comment}
\usepackage{mathtools}
\usepackage[dvipsnames]{xcolor}
\usepackage{simpler-wick}
\usepackage{multirow}
\usepackage{epstopdf}
\usepackage[utf8]{inputenc}
\usepackage[english]{babel}
\usepackage{hyperref}
\usepackage{amsfonts}
\usepackage{amssymb}
\usepackage{float}
\usepackage{hyperref}
\usepackage{braket}
\usepackage[normalem]{ulem}
\hypersetup{pdfborder=0 0 0,colorlinks=true,citecolor=blue,linkcolor=blue}
\usepackage{url}

\begin{document} 
\title{Harnessing Josephson-Shapiro physics to verify interlayer exciton superfluidity}
\author{Filippo Pascucci}
\email{filippo.pascucci@uantwerpen.be}
\affiliation{COMMIT, Department of Physics, University of Antwerp, Groenenborgerlaan 171, 2020 Antwerp, Belgium}

\author{Alexander R. Hamilton}
\affiliation{School of Physics, University of New South Wales, Sydney, New South Wales 2052, Australia}

\author{Milorad V. Milo\v{s}evi\'c}
\email{milorad.milosevic@uantwerpen.be}
\affiliation{COMMIT, Department of Physics, University of Antwerp, Groenenborgerlaan 171, 2020 Antwerp, Belgium}

\author{David Neilson}
\affiliation{COMMIT, Department of Physics, University of Antwerp, Groenenborgerlaan 171, 2020 Antwerp, Belgium}

\date{\today}


\begin{abstract}
Obtaining definitive evidence for zero-magnetic-field exciton superfluidity in electron-hole bilayers remains a longstanding challenge because the condensate is electrically neutral and its phase coherence is difficult to probe directly. We propose a direct test based on Shapiro steps in a Dayem-bridge excitonic Josephson junction. We predict clearly resolvable Shapiro plateaus in experimentally accessible current and voltage regimes for double-bilayer graphene and, in the low-density regime, double-layer transition-metal dichalcogenides. Moreover, by tuning the density across the BCS-BEC crossover we show that the Shapiro response acquires a distinct nonmonotonic evolution. This is determined by the nonmonotonic behavior of the healing length in the crossover from bosonic to fermionic excitations. Observation of these signatures would provide direct evidence of exciton superfluidity and establish exciton bilayers as a platform for neutral Josephson devices.
\end{abstract}

\maketitle

Experimental indications of excitonic superfluidity without magnetic field in bilayer electron-hole heterostructures are continuing to accumulate \cite{Burg2018, Wang2019,Gu2022, Ma2021, Nguyen2025}. Most recently, the spin-valley susceptibility in transition-metal-dichalcogenides (TMD) heterostructures showed behavior consistent with an interacting Bose-Einstein condensate (BEC) \cite{Qi2026}. 
This is an exciting possibility, since existence of such a system would establish a novel class of stable, long-lived equilibrium superfluids \cite{Lozovik1976,Ma2021,Zeng2020,Xie2018}.
By tuning the carrier densities in the device (using e.g. external metal gates), the system can be swept from strongly-coupled pair regime (BEC) with bosonic excitations to the weakly-coupled pair Bardeen-Cooper-Schrieffer (BCS) regime with fermionic excitations \cite{Pieri2007,Salasnich2013,LopezRios2018}.
Applications in electronic devices could follow, since dissipationless charge transport remains possible even for electrically neutral electron-hole pairs through counterflow in the two layers \cite{Ma2021,Su2008,Nandi2012}.

A defining feature of a superfluid is the existence of a macroscopic order parameter \cite{Ginzburg:1950}.  Unfortunately, neither spin-valley susceptibility nor  strong interlayer tunneling \cite{Spielman2000,Burg2018, Wang2019} nor perfect Coulomb drag \cite{Narozhny2016,Li2017,Liu2022,Nguyen2025}, can provide definitive evidence of the existence of such an order parameter for interlayer excitons. 
Furthermore, another standard probe of superconducting Cooper pair condensation, the Meissner effect does not occur in charge neutral exciton superfluids.  This makes it challenging to directly probe the superfluid.  

The observation of a Josephson effect would directly establish the existence of a macroscopically coherent condensate, as is performed in charged superconductors \cite{Josephson1962,Anderson1963,Sukhatme2001,Ancilotto2009,Spuntarelli2007}.  However, inserting a Josephson barrier in an exciton bilayer is not straightforward since it requires the insertion of aligned $S$-$N$-$S$ junctions in both the electron and hole layers, combining vertical stacking with lateral stitching of the 2D layers \cite{Taghinejad2018,Mahjouri-Samani2015,Frisenda2018,Gong2015}. 

Moreover, a DC Josephson effect requires resolving very small voltage drops across the junction, and for the AC effect, the Josephson frequency $f_J = \frac{2e}{h}V$ is typically very high, adding further experimental challenges. 
Using Shapiro steps converts the high-frequency Josephson dynamics into experimentally accessible DC voltage plateaus.
In this letter, we propose that Shapiro steps \cite{Shapiro1963} in a Dayem bridge Josephson junction \cite{Dayem1967, Likharev1979} could be employed as a direct phase-coherence probe of superfluidity with neutral bilayer excitons.
A Dayem bridge is a nanoconstriction joining two regions of bulk superfluid.  The current in the nanoconstriction is higher than in the bulk regions, allowing the formation of an $S$-$S'$-$S$ or $S$-$N$-$S$ junction in a single material or heterostructure without the need for additional dielectric layers.  

A notable advantage of a Josephson junction with neutral excitons is that coupling to the low-frequency \(1/f\) noise associated with charge defects and impurities would be strongly reduced \cite{Makhlin2001, Astafiev2006, Koch2007}. 
Moreover, realization of a neutral Josephson junction would open the way to a qualitatively unique class of coherent quantum devices. This includes qubits. 

The relative phase between the layers acts as a macroscopic quantum variable, analogous to the superconducting phase in conventional Josephson junctions but without charge transport. This neutral, yet electrostatically tunable, phase degree of freedom can support coherent oscillations, Shapiro locking, and phase quantization, making exciton bilayers an excellent platform for Josephson-based quantum elements.

\begin{figure}[t!]
    \centering
    \includegraphics[width=\linewidth]{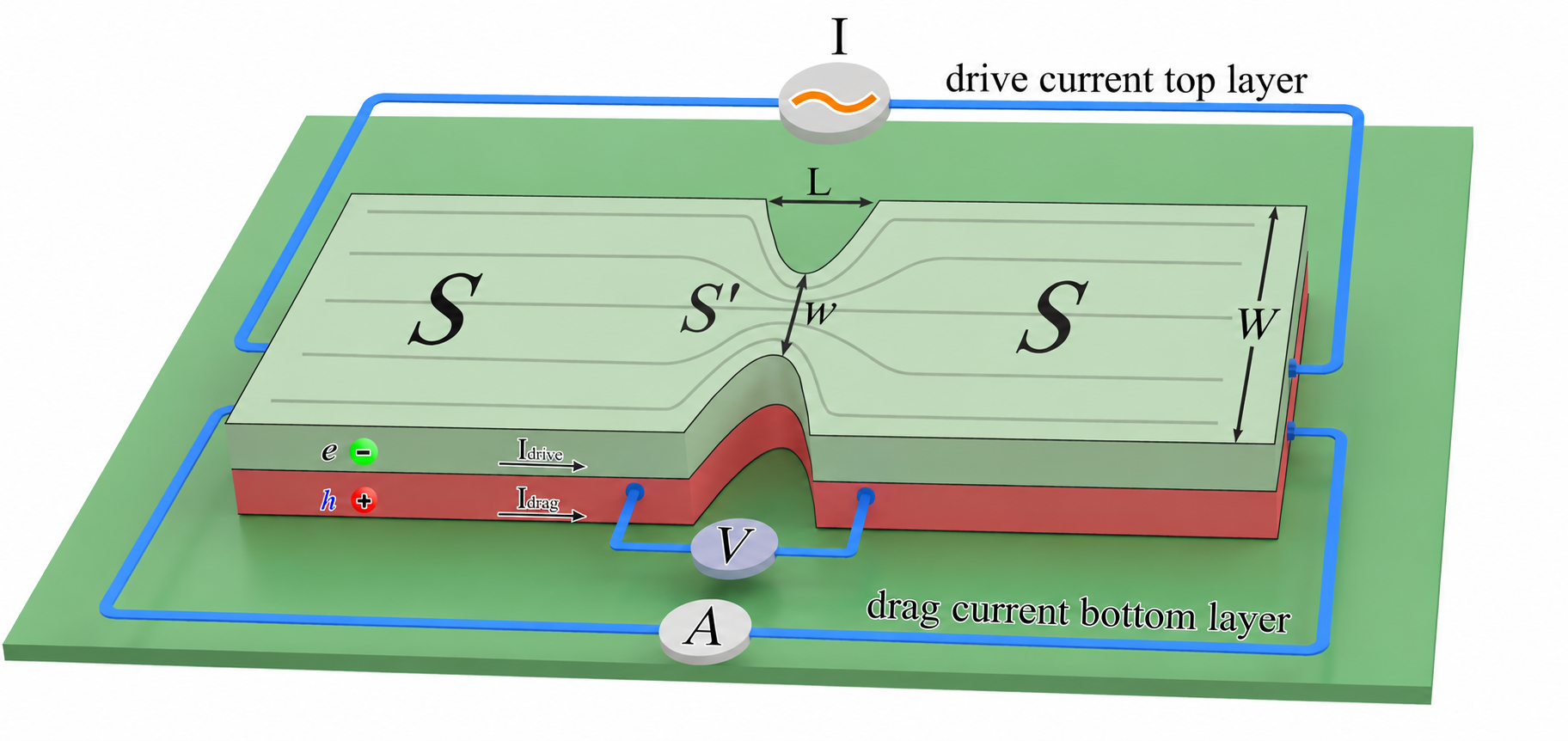}
    \caption{
    Our proposed $S$-$S'$-$S$ device and measurement setup. Top (green) layer is $n$-doped, and bottom (red) layer is $p$-doped. Bulk layer width is $W$, Dayem bridge has width $w$ and length $L$. AC current source $I$, voltage drop across the bridge $V$, ammeter $A$.
    Drive current $I_{drive}$ in the top layer induces a drag current $I_{drag}$ in the bottom layer. 
    }
    \label{Configuration}
\end{figure}

Figure \ref{Configuration} depicts our proposed Dayem bridge Josephson junction in an exciton bilayer.
We consider a pair of two-dimensional conducting layers, one $n$-doped and the other $p$-doped, electrically isolated from each other by a very thin insulator such as hexagonal-Boron-Nitride \cite{Lozovik1976, Dean2010} as in double-bilayer-graphene (DBG) system \cite{Li2016}, or by using type-II interface materials as in double-layer TMDs \cite{Gong2013}. 

In Fig.~\ref{Configuration}, the central constriction forms the Dayem bridge \cite{Dayem1967}. An injected current will result in the reduction of the superfluid order parameter in the left and right bulk leads. These are labeled in the figure as the regions $S$. The current density in the bridge  (labeled as region $S'$) will be larger than in the bulk leads, resulting in a further reduction of the order parameter in the bridge. The constriction thus acts as an $S$-$S’$-$S$ Josephson junction. Here, the weak link is created by the superflow itself rather than by an artificial barrier \cite{Gumann2007,Shelly2017,Shelly2020,Battisti2024}. In this case, although there is no Josephson tunneling, the exciton supercurrent through the Dayem bridge still follows the periodic current-phase relation analogous to the Josephson effect \cite{Gumann2007}. 

The frequency of the oscillating current in the bridge must remain below the superfluid gap energy \cite{Tinkham2004} in order to avoid driving the system into an \(S\)-\(N\)-\(S\) junction. In an \(S\)-\(N\)-\(S\) junction, formation of hotspots can produce hysteresis, excess dissipation, and voltage features that obscure the Josephson response \cite{Shelly2020}.

We propose generating the exciton current by injecting an electron current in one layer that drags holes in the adjacent layer, analogous to Coulomb drag experiments \cite{Nguyen2025}.
We consider an electric current $I=enW=ejW$ from a current source injected in the electron layer of width $W$ (Fig.\ \ref{Configuration}), where $e$ is the electron charge and $n$ is the particle density in the layer. $j=nv$ is the electron current density, where $v$ is the velocity of the electrons. 
In the superfluid region $S$ there will be perfect Coulomb drag \cite{Su2008}. Assuming equal electron and hole masses, the exciton velocity is $v_X=v/2$ because of conservation of momentum. Thus the charged currents in the layers are: 
\begin{align}
 I_{drive}= I_{e}&=env_XW=envW/2=I/2, \nonumber\\
 I_{drag}=I_{h}&=-env_XW=-I/2\ , 
\end{align}
with opposite signs as they flow in opposite directions.
The resulting neutral exciton current is $I_{X}= j_{X}W=nv_{X}W=nvW/2=I/2e$, where $j_X$ is the exciton current density. 
Within the Dayem bridge of width $w\ll W$, the current density $j_X^{\mathrm{db}}$ increases by a geometric amplification factor $\alpha=W/w\gg1$ so that    $j_X^{\mathrm{db}} =\alpha j_{X}$.

In a superfluid, the current density is related to the  gradient of the phase of the order parameter $\phi$ by,
\begin{equation}
j=\frac{\hbar n_s}{2m^*}\nabla\phi,
\label{jjosh}
\end{equation}
where $n_s$ is the superfluid density and $m^*$ the electron and hole effective masses. Imposing current conservation $I_{X}= I_X^{\mathrm{db}}$ across the Josephson junction yields,
\begin{align}
j_{X} W &= j_X^{\mathrm{db}} w, \nonumber\\
\frac{\hbar n_s W}{2m^*}\nabla\phi^{\mathrm{res}} &= \frac{\hbar n_s w}{2m^*}\nabla\phi^{\mathrm{db}}, \nonumber\\
\nabla\phi^{\mathrm{res}} &= \frac{w}{W}\nabla\phi^{\mathrm{db}} = \frac{1}{\alpha}\nabla\phi^{\mathrm{db}},
\end{align}
where \(\phi^{\mathrm{res}}\) and \(\phi^{\mathrm{db}}\) are the phase fields in the bulk reservoirs and bridge, respectively.

For $\alpha \gg 1$, the gradients of the phases 
\(\phi^{\mathrm{res}}\)  are strongly suppressed, 
leaving the phases in the left and right reservoirs, $\phi_L$ and $\phi_R$, nearly constant. To sustain the injected current, there must be a phase drop across the Dayem bridge,
\begin{equation}
\Delta\phi^{\mathrm{db}}\simeq\phi_R-\phi_L.
\label{Deltaphi}
\end{equation}
The change $\Delta\phi^{\mathrm{db}}$ will occur over the superfluid healing length $\xi_h=\hbar / (m^*v_c)$
 \cite{Gumann2007}, where $v_c$ is the critical velocity in the bridge.  
The phase gradient in $S'$ can thus be approximated by $\nabla\phi \sim {\Delta\phi^{\mathrm{db}}}/{\xi_h}$.
In the Appendix, $\xi_h$ and $v_c$ are evaluated in detail.

When the phase drop across the bridge is locked by a rigid phase difference of the reservoirs (Eq.\ \eqref{Deltaphi}), the current in the Dayem bridge can be shown to be a periodic function of $\Delta\phi^{\mathrm{db}}$ \cite{Gumann2007, Tinkham2004}. 
To maintain a single-valued current-phase relation, the junction length $L$ must be limited to $L\alt 3.5\xi_h$ \cite{Likharev1979}.

Retaining only the lowest harmonic in the expansion of the current \cite{Tinkham2004, Likharev1979}, yields  $j_X^{\mathrm{db}}=\left({\hbar n_s}/{2m^*\xi_h}\right)\sin(\Delta\phi)$, 
giving the total current in the Dayem bridge, 
\begin{equation}
I_X^{\mathrm{db}} = w\,j_X^{\mathrm{db}} = I_c^{\mathrm{JJ}}\sin(\Delta\phi),
\label{Iphi}
\end{equation}
where
\begin{equation}
I_c^{\mathrm{JJ}} = w\,\frac{\hbar n_s}{2m^*\xi_h} 
\label{Iceq}
\end{equation}
is the critical Josephson current of the Dayem bridge. Equation\ \eqref{Iphi} is the Josephson current-phase relation.

We model the bridge within the overdamped limit of the Resistive-Capacitive-Shunted-Junction  framework \cite{Russer1972,Likharev1979}, retaining the resistive channel while neglecting the capacitive contribution.
Observing Shapiro steps requires a finite voltage drop across the bridge \cite{Gross2016}. This is enabled by subgap Andreev bound states. These provide a resistive current channel in the \(S'\) region \cite{Gulubov2004,Rontani2014} and leads to time oscillations of the order parameter phase, activating the dynamical Josephson effect \cite{Tinkham2004}. 
This is the standard effective description for weakly capacitive Dayem bridges and has been shown to reproduce experimental current-voltage characteristics and Shapiro steps in superconducting \(S\)-\(S'\)-\(S\) nanobridges \cite{Shelly2017,Shelly2020}. For the neutral exciton system, the corresponding Josephson equations are, 
\begin{align}
    I_{X}&=I_c^{{\text{JJ}}} \sin(\phi(t))-\frac{2V(t)}{|e|R_n}, 
    \label{Ixbias'}\\
    \hbar\frac{d \phi(t)}{dt}&=2|e|V(t)
    \label{phit'}.
\end{align}
$R_n$ is the electric resistance of the electron layer in the normal state and $V(t)$ is the voltage drop across the Dayem bridge.
Solving Eqs.\ \eqref{Ixbias'} and \eqref{phit'} with a DC injected current $I_{X}=I_{X,DC}$ , one can explore the Josephson effects, both static ($I_{X,DC}<I_c^{{\text{JJ}}}$) and dynamic ($I_{X,DC}>I_c^{{\text{JJ}}}$). For Shapiro steps, a combined DC+AC injected current is needed, i.e. $I_{X}=I_{X,DC}+I_{X,AC}\sin(\omega_{AC} t)$.
Substituting Eq.\ \eqref{phit'} into Eq.\ \eqref{Ixbias'} gives: 
\begin{equation}
        I_{X, DC}+I_{X,AC}\sin(\omega_{AC}t)=I_c^{{\text{JJ}}} \sin(\phi(t))-\frac{\hbar}{|e|^2R_n}\frac{d\phi(t)}{dt}.
        \label{diffshap}
\end{equation}
Solving Eq.\ \eqref{diffshap} gives $\phi(t)$ and $V(t)$ as functions of $I_{X}$. 

To discuss practical results, we apply the above formalism to a Dayem bridge with width $w=50$~nm and length $L=30$~nm. 
Since typically $W\simeq10$--$20~\mu$m \cite{Burg2018, Ma2021}, the geometrical factor $\alpha\gg1$.
The condition for a single-valued current-phase relation becomes $\xi_h\gtrsim L/3.5\simeq8.5$~nm. 

\begin{figure}[t]
    \centering
    \includegraphics[width=0.48\textwidth]{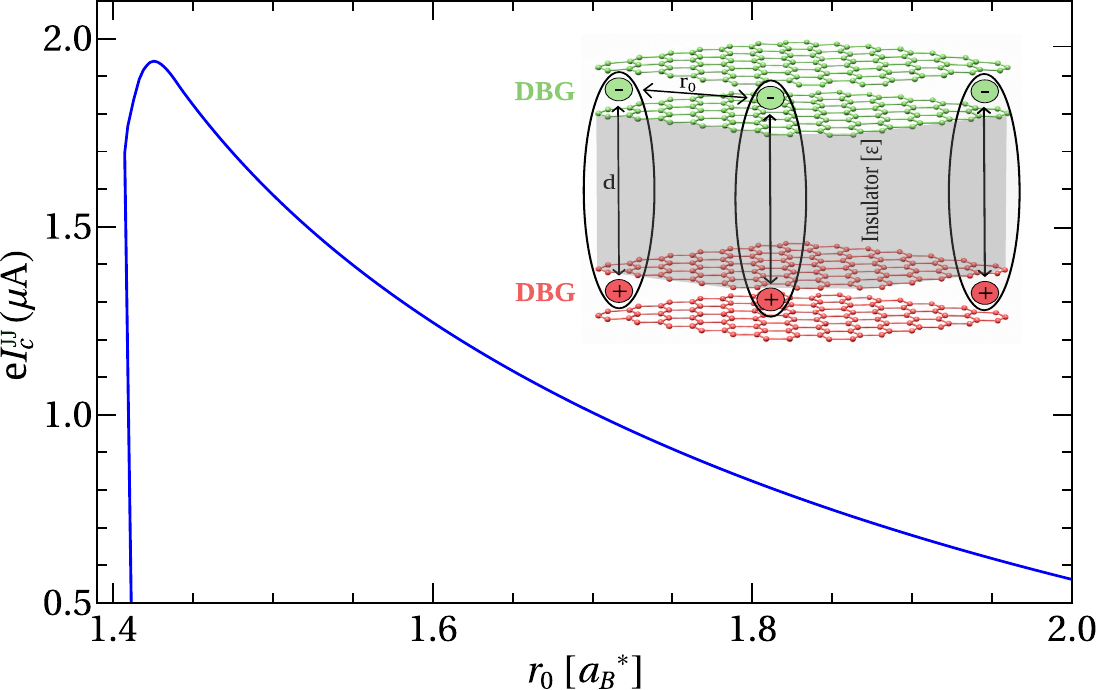}
    \caption{Critical Josephson current $eI_c^{{\text{JJ}}}$ as a function of the interparticle spacing $r_0$ for DBG. The double bilayer graphene is separated by an insulator of dielectric constant $\varepsilon$ and thickness $d$.}
    \label{Icjj}
\end{figure}

In the Appendix we determine the healing length $\xi_h$ as a function of interparticle distance $r_0$ and layer separation $d$. We consider two representative bilayer systems: Bernal stacked double bilayer-graphene and double transition-metal-dichalcogenide monolayers. 

The physical parameters are given in the Appendix. For the TMDs with layer separation $d=2$~nm, the healing length $\xi_h$ satisfies the single-value condition only at low densities, $n\alt10^{12}$~cm$^{-2}$, which is in the deep BEC regime. 
In DBG, in contrast, for $d=1$~nm the condition $\xi_h\gtrsim 8.5$~nm is satisfied across the entire range of superfluidity from BEC to the crossover regime. As detailed in the Appendix, this difference is due to the fact that the TMDs have much larger effective masses than DBG.  For this reason we focus primarily on the Shapiro steps in DBG.

We now show that the Shapiro response in exciton bilayer not only provides a direct Josephson signature, but can also resolve the evolution of the condensate across the BCS-BEC crossover. 
The key control parameter is the density dependence of the critical current $(I_c^{JJ})$, the behavior of which reflects the switch from bosonic to fermionic excitations.
\begin{figure}[t]
    \centering
    \includegraphics[width=\linewidth]{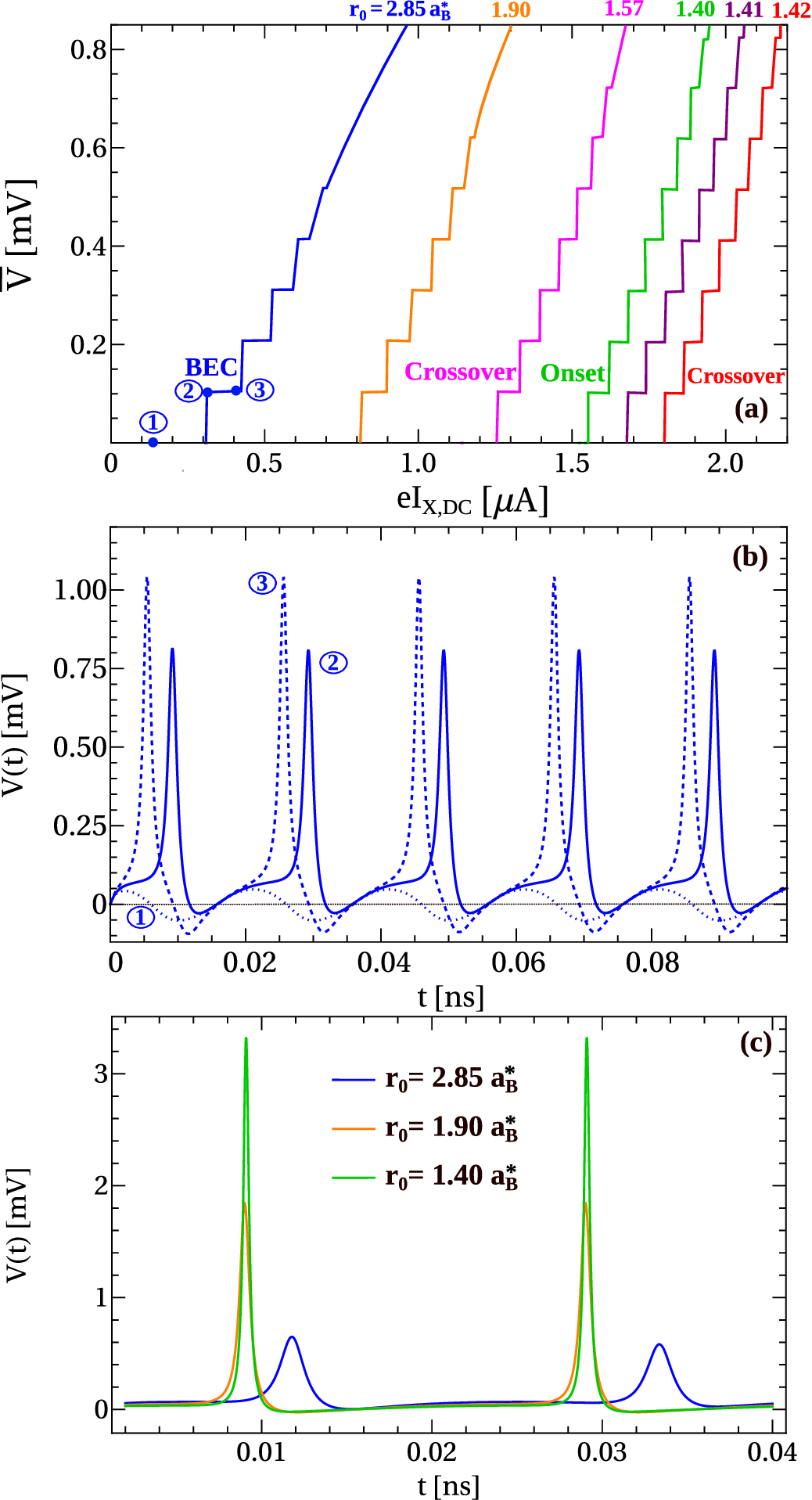}
    \caption{
    Shapiro steps in the Dayem bridge.  
    For all curves $eI_{X, AC}=0.2$ $\mu$A and $\omega_{AC}=50$ GHz. (a) I-V characteristics for different $r_0$, as labeled. $\overline{V}$ is the time voltage average and $I_{X,DC}$ is the injected exciton direct current component. \! (b) $V(t)$ at values of $eI_{X,DC}$ indicated by the circled numbers for $r_0=2.85a_B^*$. (c) $V(t)$ at the start of the first step for the values of $r_0$, as labeled.}
    \label{shapDBG}
\end{figure}
Figure~\ref{Icjj} shows the critical Josephson current $eI_c^{{\text{JJ}}}$ for DBG (Eq.\ \eqref{Iceq}). This has a maximum as a function of the interparticle spacing at $r_0\simeq1.42a_B^*$. The maximum occurs at the boundary separating the BEC and crossover regions \cite{Spuntarelli2007} and it reflects the switch from bosonic to fermionic excitations (see inset Fig.\ \ref{xih}(b) in Appendix)  \cite{Combescot2006,Pascucci2020}. The collapse of $eI_c^{{\text{JJ}}}$ at $r_{on}\simeq1.4 a_B^*$ is associated with the destruction of the superfluidity by strong screening which occurs at high densities \cite{Perali2013}. The onset density here is $n_{0}=1/\pi r_{on}^2=2.6\times10^{11}$cm$^{-2}$.

We now turn to the behavior of the Shapiro steps across the BCS-BEC crossover. Figure~\ref{shapDBG}(a) shows the $I$-$V$ characteristics $eI_{X,DC}(\overline{V})$ for fixed drive amplitude $I_{AC}=0.2\,\mu$A and frequency $\omega_{AC}=50$~GHz. We take a characteristic device resistance \(R_n \sim 1~\mathrm{k}\Omega\) typical for DBG in the low-resistance conducting regime \cite{Liu2017,Burg2018}.

The injected current interval for the condition $\overline{V}\neq 0$ follows from the behavior of the critical current (Fig.\ \ref{Icjj}), and likewise the corresponding number of steps for different $r_0$.
For this reason curves for $r_0\leq1.42\,a_B^*$ occur at decreasing injected currents respect to the $r_0\geq 1.42\,a_B^*$ ones. This is because the boundary between BEC and crossover is located at $r_0=1.42\,a_B^*$, where $I_c^{{\text{JJ}}}$ passes through a maximum.
As expected, the plateaus occur at the same voltages for all densities, $V_m=m\hbar\omega_{AC}/(2e)$, with $m$ integer. 

Figure \ref{shapDBG}(b) shows the time-domain dynamics of $V(t)$. Before the first step, $V(t)$ is sinusoidal so its time average $\overline{V}$ is zero. At the first step, $V(t)$ develops an asymmetric positive peak, making $\overline{V}=V_{m=1}>0$. Along the step as the injected DC current is increased, the peak narrows and moves to an earlier time while maintaining an unchanged oscillation frequency and an unchanged $\overline{V}$. When we jump to the next step, the profile of $V(t)$ splits into two peaks and $\overline{V}$ jumps to $V_{m=2}$. 

Figure \ref{shapDBG}(c) shows $V(t)$ at the start of the first step for three densities. 
Increasing the density sharpens the peak in $V(t)$ since the threshold current value needed to activate the first step increases with the superfluid critical current (see Fig. \ref{shapDBG}(a)). Interestingly, even within the same step, the $V(t)$ has a sensitivity to the position of the system in the BCS-BEC crossover. 

\begin{figure}[!t]
    \centering
    \includegraphics[trim=0.0cm 0.0cm 0.0cm 0.0cm, clip=true, width=0.47\textwidth]{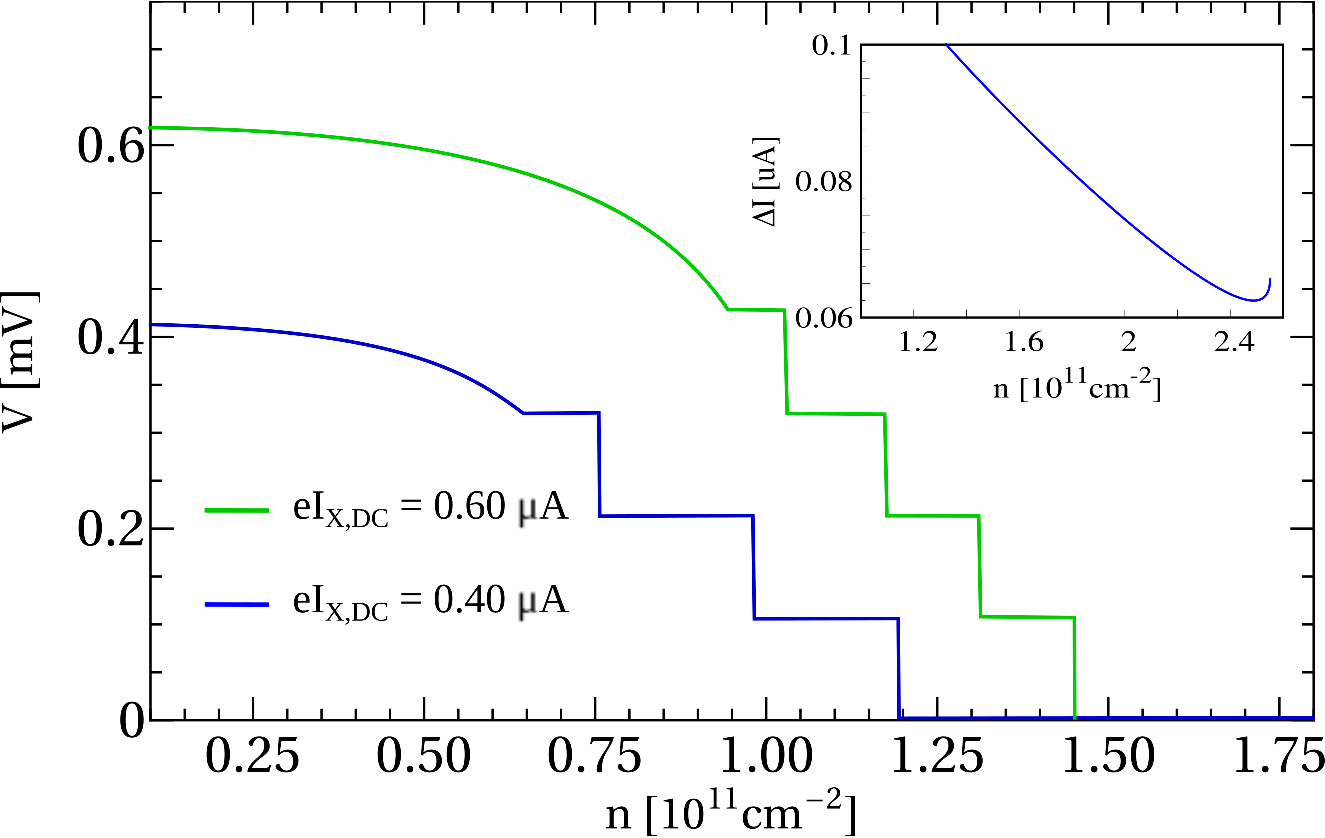}
    \caption{Voltage-density characteristics for AC current ($eI_{X,AC}=0.2\,\mu$A, $\omega_{AC}=50$~GHz), for DC currents: $eI_{X,DC}=0.4\,\mu$A (blue line) and $eI_{X,DC}=0.6\,\mu$A (green line). Inset: current width of the first step $\Delta I$ as a function of density for $eI_{X,DC}=0.4\,\mu$A.}
    \label{VIn}
\end{figure}
We demonstrate in Figure~\ref{VIn} the hitherto unexamined role of particle density, namely that it can be used as an alternative control parameter for driving the Shapiro steps, with the injected currents and frequency kept fixed.
By tuning the density, the critical superfluid current (Fig.\ \ref{Icjj}) can be controlled while sweeping across the BCS-BEC crossover, with this current the fundamental quantity regulating the Shapiro steps (Fig.\ \ref{shapDBG}).

Figure \ref{VIn} shows the density-voltage (\(n\!-\!V\)) characteristics for two values of \(I_{X,\mathrm{DC}}\).
The Shapiro steps are clearly visible across the full superfluid range. 
Increasing \(I_{X,\mathrm{DC}}\) increases $\overline{V}$, yielding more and narrower steps.
The density can thus become an alternative parameter to control the Josephson current in this system and hence the behavior of the Shapiro steps. 

We find that varying \(I_{X,\mathrm{AC}}\) and \(\omega_{\mathrm{AC}}\) produces similar effects as reported in conventional superconducting Shapiro-step measurements \cite{Gross2016}. 
Increasing \(I_{X,\mathrm{AC}}\) shifts the steps to smaller $I_{X,DC}$ and reduces their width.
Tuning \(\omega_{\mathrm{AC}}\) shifts the step voltages according to \(V_m= m\hbar\omega_{\mathrm{AC}}/2e\).

The inset in Fig.~\ref{VIn} shows that the width of the first step follows the same trend as the healing length: starting from the BEC regime and increasing the density, the width of the first step decreases, reaches a minimum at the minimum of \(\xi_h\), and then increases until the onset density. We conclude from Fig.\ \ref{shapDBG}(c) and Fig.\ \ref{VIn} that the Shapiro steps and the crossover physics are directly linked. 

In summary, we have shown that Shapiro steps in a Dayem-bridge excitonic Josephson junction provide a direct route to identifying macroscopic phase coherence in electron-hole bilayers. The Dayem-bridge geometry avoids the need for an externally fabricated tunnel barrier and realizes an \(S\)-\(S'\)-\(S\) weak link by the superflow itself. Both double-bilayer graphene and transition-metal-dichalcogenide bilayer systems can satisfy the conditions for such an \(S\)-\(S'\)-\(S\) Dayem-bridge junction. In DBG, the accessible parametric window extends across the full superfluid region from the BEC regime into the crossover regime.

We have quantified the Shapiro response not only under conventional current tuning, but also under density tuning, thereby introducing an alternative control parameter for Josephson-Shapiro physics in exciton bilayers. In this setting, the density tunes the critical Josephson current while simultaneously driving the system across the BCS-BEC crossover. This leads to previously unpredicted, clearly resolved Shapiro steps in the density-voltage characteristics at fixed injected current. Since the step structure is controlled primarily by the critical current, it should remain observable under moderate device-to-device variations in resistance and geometry.

The density dependence of the first-step width is nonmonotonic and follows the healing length, linking the dynamical Shapiro response directly to the underlying crossover physics. It reveals the change from bosonic to fermionic excitations controlling the critical current. Our results thus identify Shapiro steps as a direct probe of excitonic superfluidity and of BCS-BEC crossover physics. The same Dayem-bridge architecture may also provide a practical building block for neutral Josephson devices and, ultimately, excitonic qubit architectures with reduced low-frequency charge noise.
\\
\\
{\bf Acknowledgments}
This work was supported by Research Foundation-Flanders (FWO) (grants: 1224225N, V458124N) and by the Australian Research Council projects CE170100039, DP210101608 and IL230100072.

\newpage
\appendix

\section{End Matter}
\label{App1}

\textit{Appendix A: Healing length---}The healing length at zero temperature is defined as, 
\begin{align}
\xi_h=\frac{\hbar}{m^*v_c} \ ,
\label{xiht0}
\end{align}
with the superfluid critical velocity given by the Landau criterion \cite{Landau1941},
\begin{equation}
v_c=\min_{k}\frac{\mathcal{E}_k}{\hbar k}\ .
\label{Landaucrit}
\end{equation}
min$_{k}$ selects the $k$-value that minimizes the ratio $\mathcal{E}_k/\hbar k$.
There are two types of excitation energies $\mathcal{E}_k$ in the simplest picture, the Anderson-Bogoliubov excitations \cite{Anderson1958} associated with bosonic behavior of the pairs and the fermionic modes associated with pair-breaking excitations \cite{Combescot2006}.

In Eq.\ \eqref{Landaucrit}, $v_c$ is the lesser of the critical velocities determined for these two types of excitations.  
In the bosonic excitation branch, $\mathcal{E}_k$ is given by the dispersion relation $\mathcal{E}_k=\sqrt{\hbar^2c_{s\!f}^2k^2+\varepsilon_k^2}$ \cite{Bogoliubov1947, Combescot2006}, 
with $c_{s\!f} =\sqrt{{\mu_{s\!f}}/{2m}}$ \cite{Salasnich2015} the speed of sound. The superfluid chemical potential $\mu_{s\!f}=2\mu_s+\varepsilon_B$, $\mu_s$ is the single-particle chemical potential and $\varepsilon_B$ is the exciton binding energy.
From Eq.\ (\ref{Landaucrit}), the critical velocity for bosonic excitations is thus the speed of sound, 
\begin{eqnarray}
v_c^{({\text{BEC}})}=c_{s\!f} =\sqrt{\frac{\mu_{s\!f}}{2m}}\ .
\label{cs}
\end{eqnarray}
For the single-particle fermionic excitations, $\mathcal{E}_k=E(k)=\sqrt{\xi_k^2+\Delta_k^2}$, so the critical velocity is the pair-breaking (p-b) velocity \cite{Combescot2006},
\begin{equation}
v_c^{({\text{p-b}})}= \min_{k} \frac{\sqrt{(\varepsilon_k-\mu_s)^2+\Delta_k^2}}{\hbar k} \ . 
\label{vc(p-b)}
\end{equation}
For given values of $\mu_s$ and $\Delta_k$, 
Eq.\ \eqref{vc(p-b)} must be numerically evaluated in order 
to determine the value of $k$ that minimizes $v_c^{({\text{p-b}})}$.
The $\Delta_k$ and $\mu_s$ are obtained as functions of the density from the zero-temperature mean-field gap and number equations. Intralayer correlations are included, as are the screened electron-electron and electron-hole interactions within the Random-Phase-Approximation (see Ref.\ \cite{Pascucci2024}).

At finite temperatures, the healing length increases because of fluctuations.  In the BCS regime,  
$
    \xi_h(T)\simeq\frac{\xi_{0}}{\sqrt{1-T/T_c}}
$.  
This gives $\xi_h=1.4\xi_0$ for $T=0.5T_c$, where $T_c$ is the mean-field superfluid transition temperature.
On the other hand, in the BCS-BEC crossover and BEC regimes, 
$\xi_{h}(T)=\frac{\hbar\,v_{F}}{\pi\,\Delta(T)}$, where $\Delta(T)$ is the maximum value of the gap energy at  temperature $T$ and $v_F$ is the Fermi velocity  \cite{Tinkham2004}.

For DBG we evaluate the healing length (Eq.\ \eqref{xiht0}) for parameter values \(\varepsilon=3\varepsilon_0\), \(m^*=0.04m_e\), \(g_sg_v=4\), and \(a_B^*=7.9\,\mathrm{nm}\).  For the TMDs, we take  \(\varepsilon=5.5\varepsilon_0\), \(m^*=0.5m_e\), \(g_sg_v=2\), and \(a_B^*=1.16\,\mathrm{nm}\). For both systems we consider \(d=1\) and \(2\,\mathrm{nm}\). With \(L=30\,\mathrm{nm}\), the condition for a single-valued current-phase relation is \(\xi_h \gtrsim L/3.5 \simeq 8\,\mathrm{nm}\).

\begin{figure}[t]
    \centering
    \includegraphics[trim=0.0cm 0.0cm 0.0cm 0.0cm, clip=true, width=0.35\textwidth]{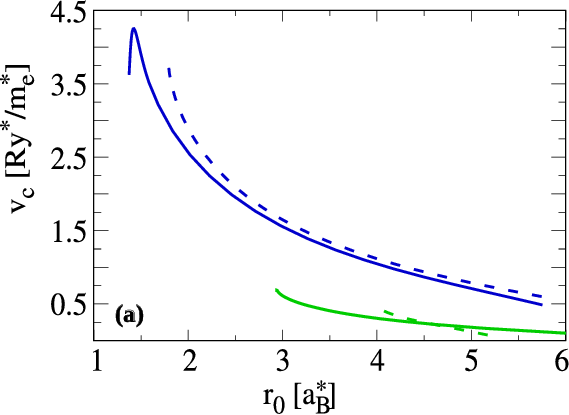}
    \includegraphics[trim=0.0cm 0.0cm 0.0cm -1.0cm, clip=true, width=0.35\textwidth]{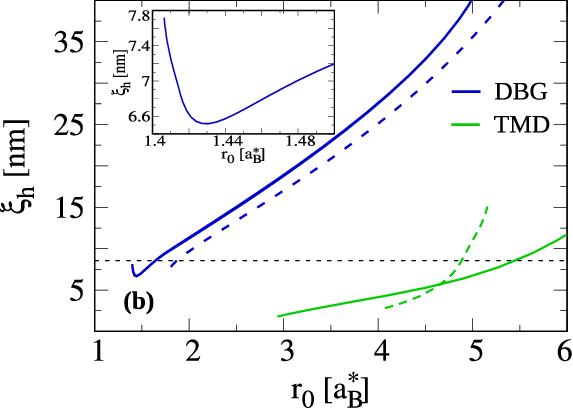}
    \caption{(a) Critical velocity $v_c$ and (b) healing length $\xi_h$ at zero temperature as functions of the average interparticle distance $r_0$. Blue lines: DBG, green lines: TMDs. Solid lines: interlayer distance $d=1$ nm, dashed lines: $d=2$ nm. In panel (b),  the black dashed line indicates the limiting condition for $S$-$S'$-$S$ at zero temperature: $[L=30$ nm$]/3.5$. Inset: zoom of minimum in healing length  near the onset density in DBG for $d=1$ nm.}
    \label{xih}
\end{figure}

The behavior of \(\xi_h\) in Fig.~\ref{xih}(b) follows directly from the critical velocity in Fig.~\ref{xih}(a). In DBG at fixed density, the \(d=1\,\mathrm{nm}\) case has a slightly stronger binding, and hence a lower chemical potential, than for \(d=2\,\mathrm{nm}\). This reduces the sound velocity [Eq.~\eqref{cs}] that sets the critical velocity over most of the density range.  This therefore gives a larger healing length. As \(d\) is reduced the onset density increases. 
For \(d=1\,\mathrm{nm}\), when increasing the density, the switch from bosonic to fermionic pair-breaking excitations occurs before the onset density. 

Interestingly, this switch is  visible in the critical velocity \cite{Pascucci2020}, resulting in an observable  maximum in \(v_c\). 
In DBG for both \(d=1\) and \(2\,\mathrm{nm}\), the condition \(\xi_h\gtrsim L/3.5\) is satisfied at finite temperatures over the full range of superfluid densities. 

In contrast, TMDs have  much shorter healing lengths since their  Coulomb interactions are stronger. This leads to  larger exciton binding energies and superfluid gaps, and an  effective Bohr radius that is nearly an order of magnitude smaller than for DBG. As a result, \(\xi_h\) is much reduced [Fig.~\ref{xih}(b)], and it exceeds (\(L/3.5\)) only at extremely low densities that lie in the deep BEC regime.

A further difference appears at low density, \(r_0>4.5\,a_B^*\), for which the TMDs system for \(d=2\,\mathrm{nm}\) has a smaller critical velocity, and hence a larger healing length, than for \(d=1\,\mathrm{nm}\). This reflects the stronger screening for \(d=2\,\mathrm{nm}\). This pushes the system  towards the crossover regime which is a less strongly interacting region. Hence at these densities the \(d=1\,\mathrm{nm}\) case remains in the BEC regime, with condensate fraction \(CF=0.88\), whereas the \(d=2\,\mathrm{nm}\) case is already in the crossover regime, with \(CF=0.75\). This is different from DBG, for which an increase of $d$ at fixed density from \(d=1\) to \(2\,\mathrm{nm}\) has very little effect on the screening, leaving the condensate fraction essentially unchanged.


\end{document}